\documentclass[twocolumn]{aastex631}
\usepackage{xspace,graphicx,apjfonts}

\newcommand\bethyi{$\beta$\,Hyi\xspace}
\newcommand\aqr{94\,Aqr\,Aa\xspace}
\newcommand{\numax}{\ensuremath{\nu_{\rm max}}}
\newcommand{\dnu}{\ensuremath{\Delta\nu}}
\newcommand{\muHz}{\mbox{$\mu$Hz}}

\shorttitle{Asteroseismology of $\beta$~Hydri}
\shortauthors{Metcalfe et al.}

\journalinfo{The Astrophysical Journal (accepted version)}

\begin{document}

\title{\Large TESS asteroseismology of $\beta$~Hydri: a subgiant with a born-again dynamo}

\author[0000-0003-4034-0416]{Travis S.~Metcalfe}
\affiliation{Center for Solar-Stellar Connections, White Dwarf Research Corporation, 9020 Brumm Trail, Golden, CO 80403, USA}

\author[0000-0002-4284-8638]{Jennifer L.~van~Saders}
\affiliation{Institute for Astronomy, University of Hawai`i, 2680 Woodlawn Drive, Honolulu, HI 96822, USA}

\author[0000-0001-8832-4488]{Daniel Huber} 
\affiliation{Institute for Astronomy, University of Hawai`i, 2680 Woodlawn Drive, Honolulu, HI 96822, USA}
\affiliation{Sydney Institute for Astronomy (SIfA), School of Physics, University of Sydney, Camperdown, NSW 2006, Australia}

\author[0000-0002-1988-143X]{Derek Buzasi}
\affiliation{Department of Chemistry and Physics, Florida Gulf Coast University, 10501 FGCU Blvd S, Fort Myers, FL 33965, USA}

\author[0000-0002-8854-3776]{Rafael A. Garc\'ia} 
\affiliation{Universit\'e Paris-Saclay, Universit\'e Paris Cit\'e, CEA, CNRS, AIM, 91191, Gif-sur-Yvette, France}

\author[0000-0002-3481-9052]{Keivan G.~Stassun}
\affiliation{Department of Physics and Astronomy, Vanderbilt University, Nashville, TN 37235, USA}

\author[0000-0002-6163-3472]{Sarbani Basu}
\affiliation{Department of Astronomy, Yale University, PO Box 208101, New Haven, CT 06520-8101, USA}

\author[0000-0003-0377-0740]{Sylvain N.~Breton}
\affiliation{INAF – Osservatorio Astrofisico di Catania, Via S. Sofia, 78, 95123 Catania, Italy}

\author[0000-0002-9879-3904]{Zachary R.~Claytor}
\affiliation{Department of Astronomy, University of Florida, 211 Bryant Space Science Center, Gainesville, FL 32611, USA}
\affiliation{Space Telescope Science Institute, 3700 San Martin Drive, Baltimore, MD 21218, USA}

\author[0000-0001-8835-2075]{Enrico Corsaro}
\affiliation{INAF – Osservatorio Astrofisico di Catania, Via S. Sofia, 78, 95123 Catania, Italy}

\author[0000-0001-9169-2599]{Martin B.~Nielsen}
\affiliation{School of Physics \& Astronomy, University of Birmingham, Edgbaston, Birmingham B15 2TT, UK}

\author[0000-0001-7664-648X]{J.~M.~Joel Ong}
\altaffiliation{NASA Hubble Fellow}
\affiliation{Institute for Astronomy, University of Hawai`i, 2680 Woodlawn Drive, Honolulu, HI 96822, USA}

\author[0000-0003-2657-3889]{Nicholas Saunders}
\altaffiliation{NSF Graduate Research Fellow}
\affiliation{Institute for Astronomy, University of Hawai`i, 2680 Woodlawn Drive, Honolulu, HI 96822, USA}

\author[0000-0002-5496-365X]{Amalie Stokholm}
\affiliation{School of Physics \& Astronomy, University of Birmingham, Edgbaston, Birmingham B15 2TT, UK}
\affiliation{Stellar Astrophysics Centre, Aarhus University, Ny Munkegade 120, DK-8000 Aarhus C, Denmark}

\author[0000-0001-5222-4661]{Timothy R.~Bedding}
\affiliation{Sydney Institute for Astronomy (SIfA), School of Physics, University of Sydney, Camperdown, NSW 2006, Australia}

\begin{abstract}

The solar-type subgiant \bethyi has long been studied as an old analog of the Sun. 
Although the rotation period has never been measured directly, it was estimated to be 
near 27~days. As a southern hemisphere target it was not monitored by long-term stellar 
activity surveys, but archival International Ultraviolet Explorer data revealed a 12~year 
activity cycle. Previous ground-based asteroseismology suggested that the star is 
slightly more massive and substantially larger and older than the Sun, so the similarity 
of both the rotation rate and the activity cycle period to solar values is perplexing. We 
use two months of precise time-series photometry from the Transiting Exoplanet Survey 
Satellite (TESS) to detect solar-like oscillations in \bethyi and determine the 
fundamental stellar properties from asteroseismic modeling. We also obtain a direct 
measurement of the rotation period, which was previously estimated from an ultraviolet 
activity-rotation relation. We then use rotational evolution modeling to predict the 
rotation period expected from either standard spin-down or weakened magnetic braking 
(WMB). We conclude that the rotation period of \bethyi is consistent with WMB, and that 
changes in stellar structure on the subgiant branch can reinvigorate the large-scale 
dynamo and briefly sustain magnetic activity cycles. Our results support the existence of 
a ``born-again'' dynamo in evolved subgiants---previously suggested to explain the cycle 
in \aqr---which can best be understood within the WMB scenario.

\end{abstract}

%\keywords{Stellar activity; Stellar evolution; Stellar oscillations; Stellar rotation}

%%%%%%%%%%%%%%%%%%%%%%%%%%%%%%%%%%%%%%%%%%%%%%%%%%%%%%%%%%%%%%%%%%%%%%%%%%%%%%%%%%%%%%%%%% 
\section{Introduction}\label{sec1}

The solar-type subgiant \bethyi (HD\,2151, TIC\,267211065) has been studied for decades 
as an old analog of the Sun \citep{Dravins1993a, Dravins1993b, Dravins1993c, 
Dravins1998}. Some of the earliest attempts to detect solar-like oscillations only 
produced upper limits \citep{Frandsen1987, Edmonds1995}, but high-precision radial 
velocity measurements ultimately identified excess power around 1~mHz \citep{Bedding2001, 
Carrier2001} and a subsequent dual-site campaign resolved the individual mode frequencies 
\citep{Bedding2007}. These observations led to some initial asteroseismic modeling of the 
global properties \citep{DiMauro2003, Fernandes2003} followed by more precise modeling of 
the individual oscillation frequencies \citep{Dogan2010, Brandao2011}, revealing a star 
that is slightly more massive and substantially larger and older than the Sun.

\begin{figure}[t]
 \centering\includegraphics[width=\columnwidth]{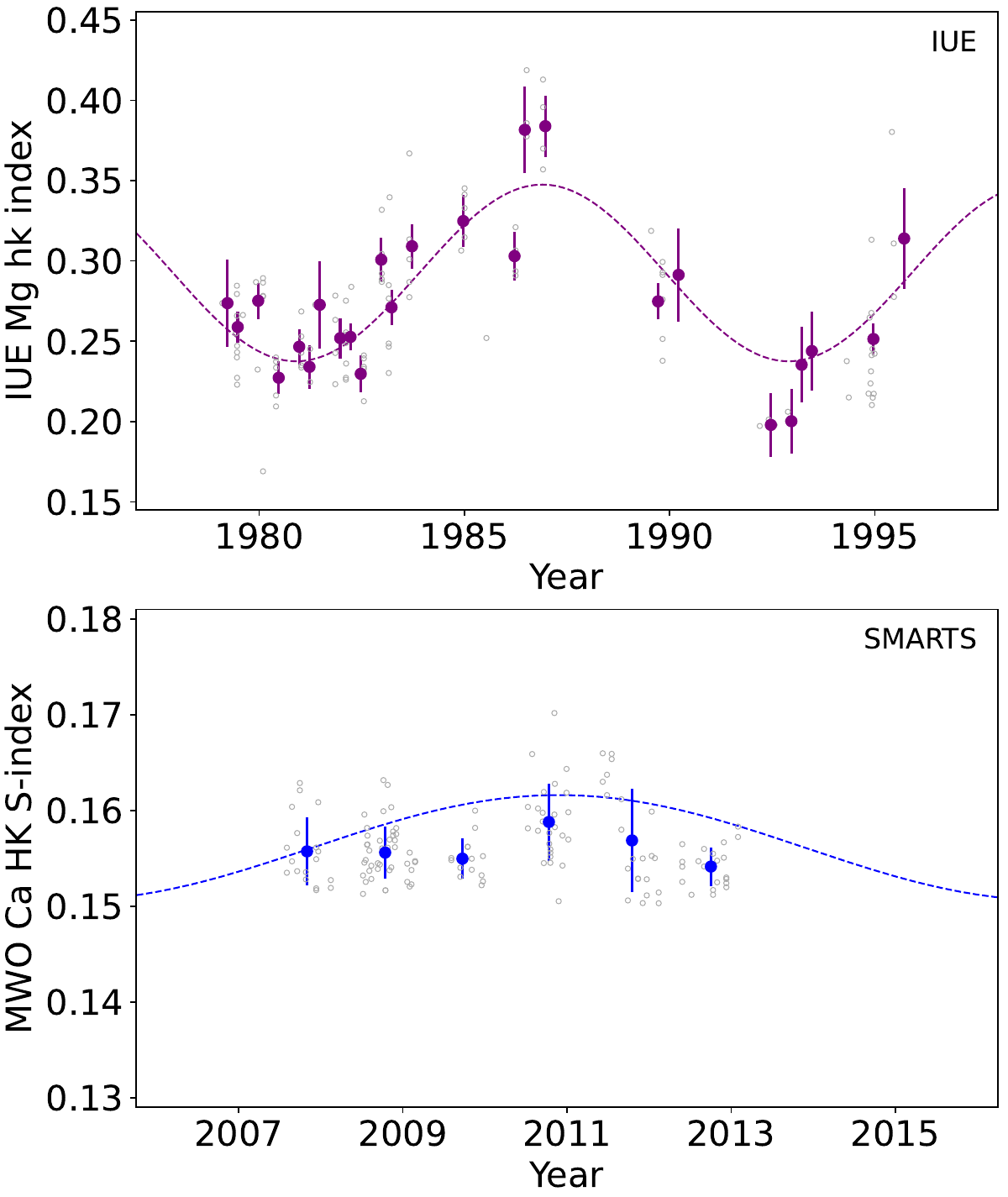}
 \caption{Activity cycle of \bethyi from IUE data \citep[top, adapted 
from][]{Metcalfe2007} and from ground-based observations (bottom). The horizontal and 
vertical scales of the two panels differ by factors of two and six, respectively. 
Fractional variability due to the cycle is about six times larger in the ultraviolet than 
in the optical.\label{fig1}}
 \end{figure}

As a southern hemisphere target \bethyi was not monitored by the Mount Wilson survey, but 
archival International Ultraviolet Explorer (IUE) data revealed a 12\,yr activity cycle 
\citep[][see Figure~\ref{fig1}]{Metcalfe2007}. These same IUE observations were later 
used to estimate the rotation period ($P_{\rm rot}$), adopting an ultraviolet 
activity-rotation relation established from a larger sample of stars \citep[$P_{\rm rot} 
= 27.1 \pm 1.7$~days;][]{Olmedo2013}. Activity cycles are rare in subgiants, with most of 
the evolved stars in the Mount Wilson survey showing constant activity over decades 
\citep{Baliunas1995}. The only two unambiguous examples of subgiants with cycling 
activity include HD\,81809 \citep{Egeland2018} and \aqr 
\citep[HD\,219834A;][]{Metcalfe2020}, which are both members of binary or multiple star 
systems. The rotation and cycle period ($P_{\rm cyc}$) of HD\,81809 places it near the 
short-period sequence of stellar activity cycles \citep{BohmVitense2007, 
Brandenburg2017}, while both \aqr and \bethyi appear to be outliers in a $P_{\rm rot}$ 
versus $P_{\rm cyc}$ diagram.

% FIGURE 1 ---------------------------------------------------------------   
%-------------------------------------------------------------------------               

We use recent observations from the Transiting Exoplanet Survey Satellite 
\citep[TESS;][]{Ricker2014} to constrain the global properties of \bethyi and investigate 
the possible origins of its activity cycle. In Section~\ref{sec2} we describe the TESS 
observations, and we derive new constraints on both the rotation period and the stellar 
luminosity. In Section~\ref{sec3} we analyze the TESS data to extract the solar-like 
oscillation frequencies, and we determine the fundamental stellar properties from 
asteroseismic modeling. In Section~\ref{sec4} we use rotational evolution models to probe 
standard spin-down and weakened magnetic braking (WMB) scenarios, and we propose that 
changes in stellar structure on the subgiant branch can reinvigorate the large-scale 
dynamo and briefly sustain magnetic activity cycles. Finally, in Section~\ref{sec5} we 
summarize and discuss our results, concluding that ``born-again'' dynamos in evolved 
subgiants can best be understood within the WMB scenario.

%%%%%%%%%%%%%%%%%%%%%%%%%%%%%%%%%%%%%%%%%%%%%%%%%%%%%%%%%%%%%%%%%%%%%%%%%%%%%%%%%%%%%%%%%% 
\section{Observations}\label{sec2}

\subsection{TESS Photometry}\label{sec2.1}

TESS observed \bethyi at 20~s cadence during Sectors 67 and 68, corresponding to 2023 
July 1 -- 2023 August 25, with the usual gaps mid-sector and between sectors for data 
downlink. We used 20~s instead of 2~min data due to the significant improvement in 
photometric precision for bright stars \citep{Huber2022}. Given the brightness of 
\bethyi (${\rm Tmag} = 2.216$), we anticipated count rates of approximately $2 \times 
10^7~\rm s^{-1}$, which is consistent with the rates seen in the Science Processing 
Operations Center \citep[SPOC;][]{Jenkins2016} simple aperture photometry (SAP) light 
curves for both sectors. However, the corrected SPOC Pre-search Data Conditioning SAP 
(PDCSAP) light curve for both sectors shows a mean value of approximately $1.14 \times 
10^7~\rm s^{-1}$, substantially lower than both expectations and the uncorrected version. 
Both SPOC light curves also show noise levels that are significantly higher than 
anticipated for such a bright star.

% FIGURE 2 ---------------------------------------------------------------   
 \begin{figure}[t]
 \centering\includegraphics[width=\columnwidth]{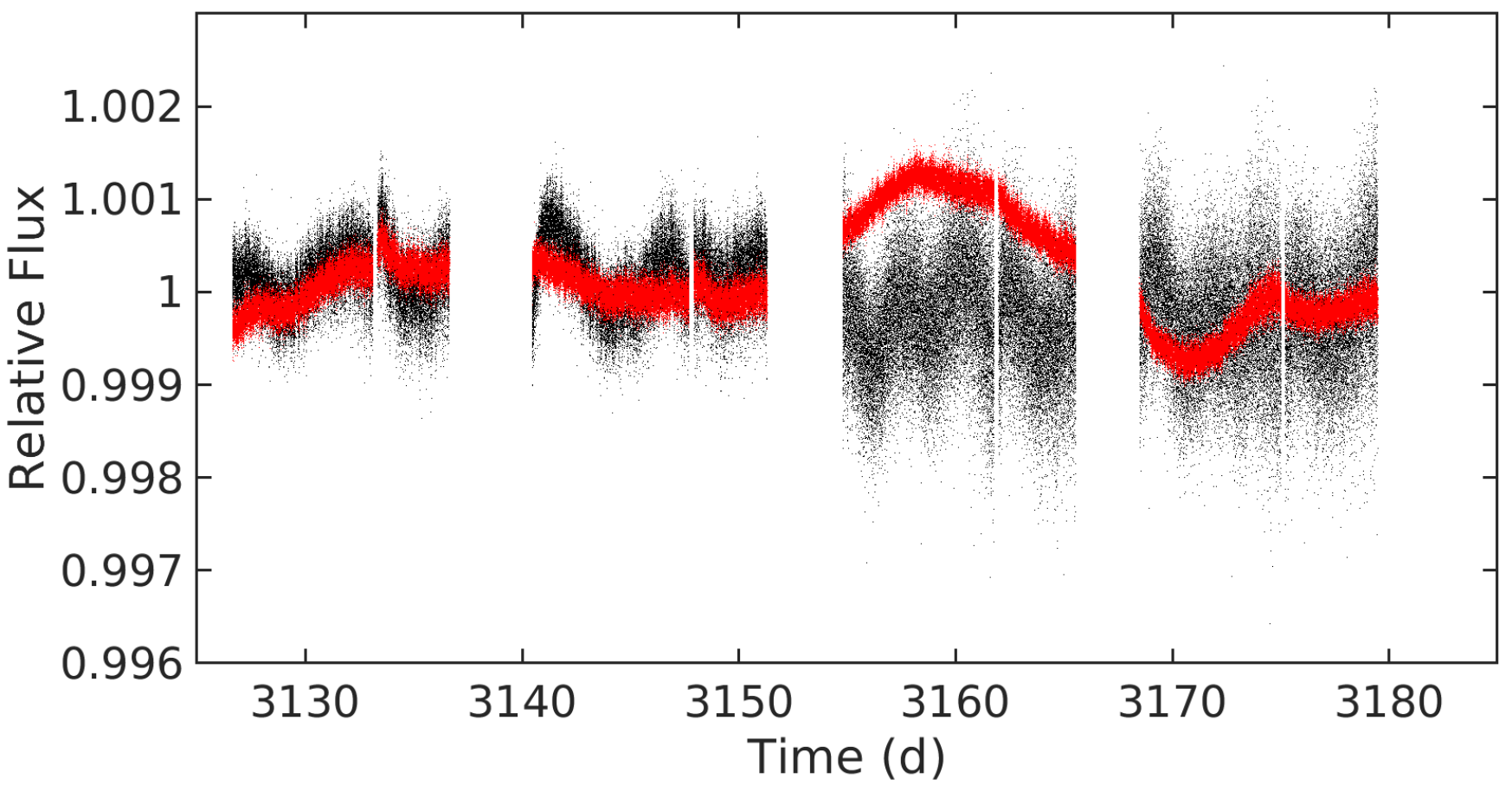}
 \caption{Light curves for \bethyi from TESS data. Red points show the unfiltered light 
curve described in Section~\ref{sec2.1}, compared to the PDCSAP light curve in black. Our 
procedure reduces the {\it rms} noise level of the light curve by approximately a factor 
of two.\label{fig2}}
 \end{figure}
%-------------------------------------------------------------------------               

From experience, we know that the normal TESS data processing can struggle with bright 
sources such as \bethyi. Accordingly, we followed the procedures described in 
\cite{Nielsen2020} and \cite{Metcalfe2023b} to extract custom light curves for each 
sector. In essence, our aperture masks are created by starting from the pixel with the 
largest flux, and adding additional pixels one at a time until the signal-to-noise ratio 
(S/N) no longer improves. The resulting light curve is then detrended against centroid 
pixel coordinates and high-pass filtered with a cutoff frequency of $100~\mu \rm Hz$ to 
minimize residual contributions from spacecraft positional jitter. Figure~\ref{fig2} 
compares our unfiltered light curve (red) to the \mbox{PDCSAP} data product (black), 
showing the approximately factor of two reduction in the root-mean-square ({\it rms}) 
noise level of the light curve.

\subsection{Signatures of Rotation}\label{sec2.2}

To look for any signature of rotation in the unfiltered light curve described in 
Section~\ref{sec2.1}, we started by interpolating the gaps with inpainting techniques 
based in a sparsity prior \citep{Elad05} using a multiscale discrete cosine transform 
\citep{Starck2006, Garcia2014a, Pires2015}. This procedure has been successfully applied 
to determine the surface rotation periods of main-sequence stars, subgiants, and red 
giants \citep[e.g.][]{Santos2019, Santos2021, Lam2021, Garcia2023}. The resulting light 
curve was re-binned by 100 points to a final cadence of $\sim$33~minutes.

% FIGURE 3 ---------------------------------------------------------------   
 \begin{figure}[t]
 \centering\includegraphics[width=0.97\columnwidth, trim=3.5cm 4.1cm 2.8cm 2.3cm, clip]{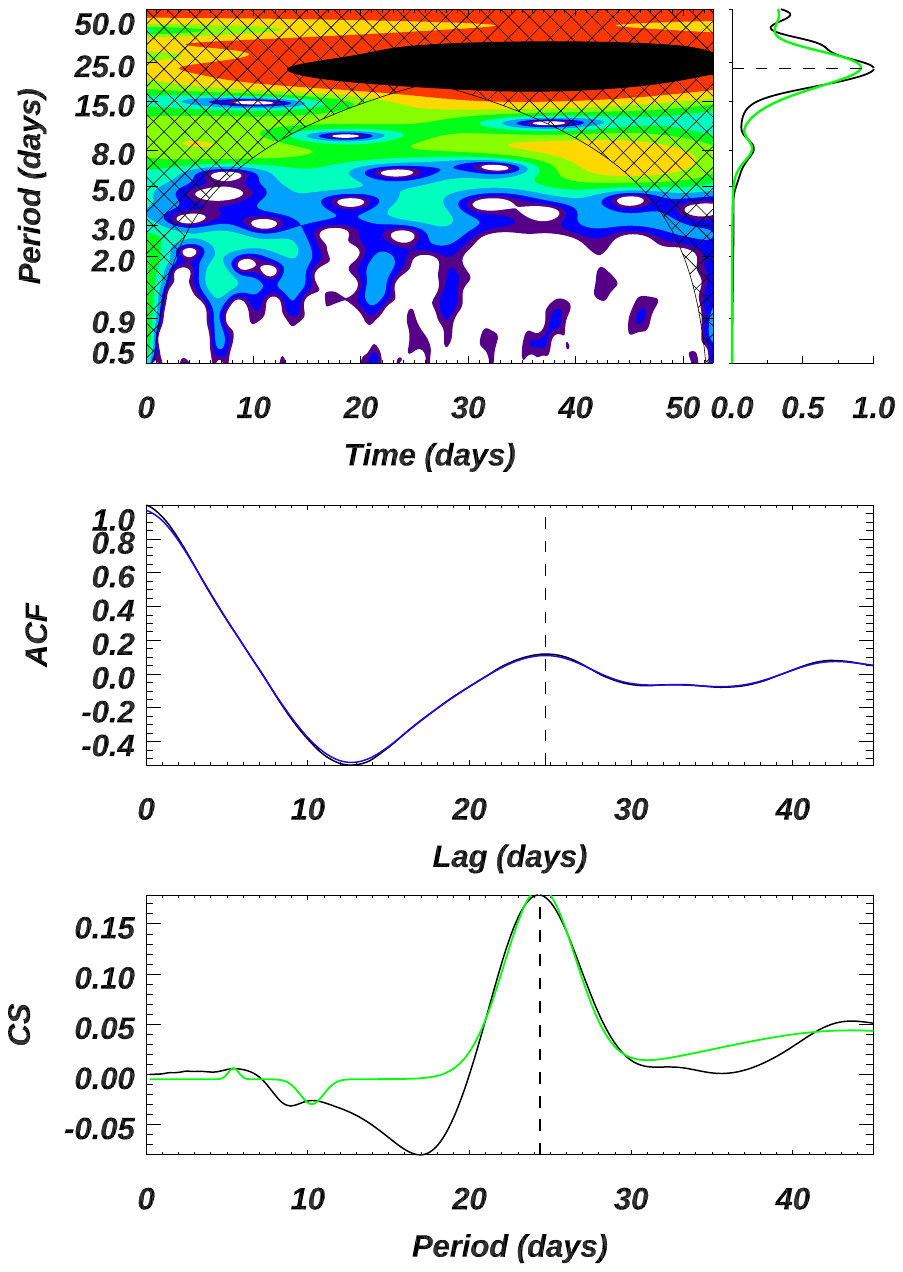}
 \caption{Analysis of the TESS light curve to determine $P_{\rm{rot}}$, including a 
period-time analysis (top left) projected onto the period axis and normalized by the 
maximum power (top right), an auto-correlation analysis (middle), and the composite 
spectrum (bottom). The hashed region in the top panel corresponds to the area where the 
retrieved periods are less reliable. Green lines correspond to the Gaussian fits (see 
Section~\ref{sec2.2} for details).\label{fig3}}
 \end{figure}
%------------------------------------------------------------------------- 

The rotation period estimate was obtained from three different but complementary methods, 
which are illustrated in Figure~\ref{fig3}. The initial method involved a period-time 
analysis (top left panel), using a Morlet wavelet-based decomposition 
\citep{TorrenceCampo1998}. We projected the results onto the period axis, creating the 
global wavelet power spectrum \citep[GWPS;][]{Mathur2010, Garcia2014b}. Starting from the 
highest peak, we iteratively fit Gaussian functions (green line, top right panel) and 
subtracted them from the GWPS until no further power remained. The central value of the 
tallest Gaussian function serves as our initial estimate of $P_{\rm{rot}}$. We adopt the 
full width at half maximum as the measurement uncertainty---a conservative approach that 
considers the possibility of surface latitudinal differential rotation. A second estimate 
of $P_{\rm{rot}}$ was obtained from the auto-correlation function 
\citep[ACF;][]{McQuillan2013} of the light curve (middle panel). Finally, we multiplied 
the GWPS and the ACF to obtain the composite spectrum \citep[CS;][]{Ceillier2016, 
Ceillier2017}, which emphasizes signals common to both analyses (bottom panel).
 
After bench-marking several different surface rotation pipelines, \cite{Aigrain2015} 
showed that combining different analysis techniques is a powerful method to determine a 
reliable value of $P_{\rm{rot}}$. Starting with the GWPS, the main periodicity 
corresponds to $\sim$23 days. Unfortunately, this value falls within the exclusion cone 
of the wavelet analysis, indicated by the hashed region in the top left panel of 
Figure~\ref{fig3}. Reliable results typically require a light curve longer than three 
times the rotation period. The ACF shows a signal around 24.7 days, which is further 
enhanced in the CS (24.4 days). Given the star's brightness, we believe that the 
modulation is likely of stellar origin. Based on the current TESS data analysis, we 
conclude that the rotation period is $P_{\rm rot}=23.0 \pm 2.8$~days. However, additional 
observations may be required to confirm this estimate. Note that previous single-sector 
data do not provide useful constraints.

\subsection{Spectral Energy Distribution}\label{sec2.3}

To obtain a constraint on the stellar luminosity, we performed an analysis of the 
broadband spectral energy distribution (SED) of \bethyi together with the {\it Gaia} DR3 
parallax, following the procedures described in \cite{Stassun2016} and \cite{Stassun2017, 
Stassun2018}. No systematic offset was applied \citep[see, e.g.,][]{StassunTorres2021}. 
We adopted the $JHK_S$ magnitudes from {\it 2MASS}, the W3--W4 magnitudes from {\it 
WISE}, the $UBV$ magnitudes from \cite{Mermilliod2006}, and the Str\"omgren $ubvy$ 
magnitudes from \cite{Paunzen2015}. Together, the available photometry spans the full 
stellar SED over the wavelength range 0.4--20~$\mu$m.

We performed a fit using Kurucz stellar atmosphere models, with the effective temperature 
($T_{\rm eff}$) from \cite{North2007}, and the surface gravity ($\log g$) and metallicity 
([Fe/H]) from \cite{Bruntt2010}. The extinction $A_V$ was fixed at zero due to the very 
close proximity of the star ($d=7.5$~pc). The resulting fit has a reduced $\chi^2$ of 
0.9. Integrating the model SED gives the bolometric flux at Earth, $F_{\rm bol} = 1.979 
\pm 0.059 \times 10^{-6}$ erg~s$^{-1}$~cm$^{-2}$. Taking $F_{\rm bol}$ together with the 
{\it Gaia} parallax \citep{Gaia2021} yields the bolometric luminosity $L_{\rm bol} = 3.45 
\pm 0.10$~L$_\odot$.

%%%%%%%%%%%%%%%%%%%%%%%%%%%%%%%%%%%%%%%%%%%%%%%%%%%%%%%%%%%%%%%%%%%%%%%%%%%%%%%%%%%%%%%%%% 
\section{Asteroseismology}\label{sec3}

\subsection{Solar-like Oscillation Frequencies}\label{sec3.1}

\begin{figure}
 \centering\includegraphics[width=\columnwidth,trim=15 15 10 10,clip]{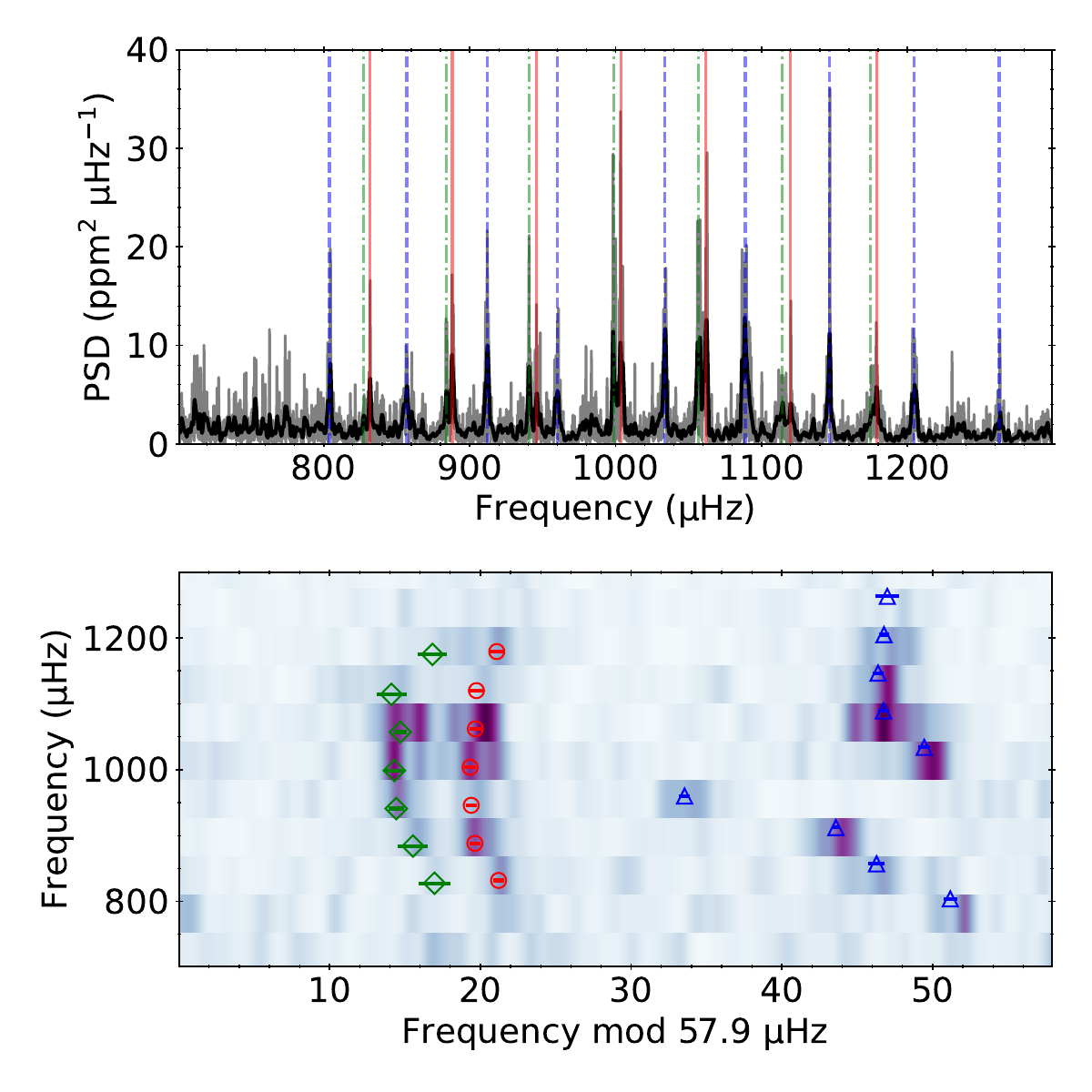} 
\caption{Power spectrum (top) and \'{e}chelle diagram (bottom) centered on the power 
excess due to solar-like oscillations detected in \bethyi. Red solid lines and circles 
indicate extracted radial ($l=0$) modes, blue dashed lines and squares show extracted 
dipole ($l=1$) modes, and green dash-dotted lines and diamonds show extracted quadrupole 
($l=2$) modes.\label{fig4}}
 \end{figure}

The power spectrum of the filtered TESS light curve for \bethyi is shown in the top panel 
of Figure~\ref{fig4}. We observe a clear series of regularly spaced peaks centered at a 
frequency near \numax=1038~\muHz, consistent with previous detections of solar-like 
oscillations. To extract individual frequencies, three teams applied Lorentzian 
mode-profile fitting \citep[e.g.][]{Garcia2009, Handberg2011, Appourchaux2012, 
Mosser2011c, Corsaro2014, Corsaro2020, Nielsen2021, Breton2022}. For each mode, we 
required two independent methods to return the same frequency within uncertainties. The 
consensus list of frequencies was further refined from visual inspection of the power 
spectrum. For the final list we adopted values from a single method, with uncertainties 
derived by adding in quadrature the median formal uncertainty and the standard deviation 
of the extracted frequencies from all methods that identified a given mode. 
Table~\ref{tab1} lists the identified frequencies, which agree with the corresponding 
modes from \cite{Bedding2007} within the uncertainties.

% FIGURE 4 ---------------------------------------------------------------   
%-------------------------------------------------------------------------

The bottom panel of Figure~\ref{fig4} shows an \'{e}chelle diagram with a large 
separation of $\dnu=57.9$~\muHz\ and the extracted frequencies. We identified seven 
radial ($l=0$), nine dipole ($l=1$), and seven quadrupole ($l=2$) modes. The dipole 
modes show a clear avoided crossing near 950~\muHz, a departure from the regular 
frequency spacing due to the interaction of gravity modes in the core and pressure modes 
in the envelope \citep{Aizenman1977}. Avoided crossings are often found in the 
oscillations of subgiants, and are a powerful diagnostic of the stellar age 
\citep[e.g.][]{Benomar2012}.

% TABLE 1 ----------------------------------------------------------------   
\begin{deluxetable}{ccc}[t]
  \setlength{\tabcolsep}{28pt}
  \tablecaption{Identified Oscillation Frequencies for \bethyi \label{tab1}}
  \tablehead{\colhead{$\nu~(\muHz)$} & \colhead{$\sigma_{\nu}~(\muHz)$} & \colhead{$l$}}
  \startdata
  803.86  & 0.46 & 1 \\
  827.55  & 1.06 & 2 \\
  831.80  & 0.36 & 0 \\
  856.87  & 0.54 & 1 \\
  884.02  & 1.00 & 2 \\
  888.13  & 0.35 & 0 \\
  912.08  & 0.29 & 1 \\
  940.81  & 0.54 & 2 \\
  945.79  & 0.36 & 0 \\
  959.94  & 0.37 & 1 \\
  998.59  & 0.61 & 2 \\
  1003.61 & 0.37 & 0 \\
  1033.74 & 0.37 & 1 \\
  1056.88 & 0.44 & 2 \\
  1061.86 & 0.37 & 0 \\
  1088.93 & 0.35 & 1 \\
  1114.19 & 0.99 & 2 \\
  1119.83 & 0.50 & 0 \\
  1146.47 & 0.37 & 1 \\
  1174.82 & 0.97 & 2 \\
  1179.08 & 0.45 & 0 \\
  1204.77 & 0.35 & 1 \\
  1262.89 & 0.77 & 1 \\
  \enddata
\end{deluxetable}
\vspace*{-12pt}
%-------------------------------------------------------------------------

\subsection{Asteroseismic Modeling}\label{sec3.2}

To determine the fundamental properties of \bethyi, several modeling teams attempted to 
match the oscillation frequencies identified in Section~\ref{sec3.1}, along with the 
$T_{\rm eff}$ from \cite{North2007} and the [Fe/H] from \cite{Bruntt2010} with errors 
inflated to account for the systematic noise floor suggested by \cite{Torres2012}, as 
well as the luminosity constraint from Section~\ref{sec2.3}. The five independent 
analyses used stellar evolution models from ASTEC \citep{ChristensenDalsgaard2008a}, 
GARSTEC \citep{Aguirre2022}, MESA \citep{Paxton2011}, and YREC \citep{Demarque2008}. The 
resulting determinations of the stellar properties were in very good agreement, with a 
relative dispersion of 2\% in radius (1.796--1.840~$R_\odot$), 6\% in mass 
(1.059--1.127~$M_\odot$), and 22\% in age (6.1--7.5~Gyr). For consistency with the 
rotational evolution modeling in Section~\ref{sec4.1}, below we provide additional 
details about the results from YREC and MESA \citep[AMP~2.0;][]{AMP2023}.

% TABLE 2 ----------------------------------------------------------------   
\begin{deluxetable}{@{}lcc}[t]
  \setlength{\tabcolsep}{16pt}
  \tablecaption{Properties of the Solar-type Subgiant \bethyi \label{tab2}}
  \tablehead{\colhead{}                & \colhead{\bethyi}      & \colhead{Source}}
  \startdata
  $T_{\rm eff}$ (K)                    & $5872 \pm 74$          & (1) \\
  $[$Fe/H$]$ (dex)                     & $-0.10 \pm 0.09$       & (2) \\
  $\log g$ (dex)                       & $3.84 \pm 0.08$        & (2) \\
  $B-V$ (mag)                          & $0.618$                & (3) \\
  $\log R'_{\rm HK}$ (dex)             & $-4.996$               & (3) \\
  $P_{\rm cyc}$ (yr)                   & $12.0^{+3.0}_{-1.7}$   & (4) \\
  $P_{\rm rot}$ (days)                 & $23.0 \pm 2.8$         & (5) \\ 
  Luminosity ($L_\odot$)               & $3.45 \pm 0.10$        & (6) \\
  Radius ($R_\odot$)~~[YREC]           & $1.831 \pm 0.009$      & (7) \\ 
  \phantom{Radius ($R_\odot$)}~~[MESA] & $1.840 \pm 0.032$      & (7) \\ 
  Mass ($M_\odot$)~~[YREC]             & $1.107 \pm 0.009$      & (7) \\ 
  \phantom{Mass ($M_\odot$)}~~[MESA]   & $1.127 \pm 0.054$      & (7) \\ 
  Age (Gyr)~~[YREC]                    & $6.46 \pm 0.13$        & (7) \\ 
  \phantom{Age (Gyr)}~~[MESA]          & $6.26 \pm 0.57$        & (7) \\ 
  \enddata
  \tablerefs{(1)~\cite{North2007}; (2)~\cite{Bruntt2010}; (3)~\cite{Henry1996}; 
  (4)~\cite{Metcalfe2007}; (5)~Section\,\ref{sec2.2}; (6)~Section\,\ref{sec2.3}; 
  (7)~Section\,\ref{sec3.2}}
\vspace*{-24pt}
\end{deluxetable}
\vspace*{-12pt}
%-------------------------------------------------------------------------                             

The asteroseismic modeling with YREC generally followed the same procedures described in 
\cite{Metcalfe2020} for the subgiant \aqr. However, the characteristics of the model 
grids were different for \bethyi, and the treatment of spectroscopic constraints was 
modified slightly. The initial model grid for \bethyi was constructed with masses between 
1.02 and 1.20~$M_\odot$ in increments of 0.01~$M_\odot$, initial helium abundances 
between 0.25 and 0.29, initial [Fe/H] between $-$0.15 and +0.15, and a mixing length 
parameter between 1.4 and 2.1. After evaluating the most likely mass, a finer grid was 
constructed with masses between 1.09 and 1.12~$M_\odot$ and increments of 
0.005~$M_\odot$, with all other parameters spanning the same ranges as in the initial 
grid. This finer grid was then used for a Monte Carlo analysis of the spectroscopic 
parameters. For each realization of the spectroscopic constraints, the most likely 
asteroseismic model was identified. The final stellar properties were obtained from the 
distribution of most likely models resulting from this procedure.

The MESA results were obtained from version 2.0 of the Asteroseismic Modeling Portal 
(AMP)\footnote{\url{https://github.com/travismetcalfe/AMP2}}, which originally used 
models from the Aarhus stellar evolution and pulsation codes 
\citep{ChristensenDalsgaard2008a, ChristensenDalsgaard2008b}. The AMP code was released 
in 2009 \citep{Metcalfe2009, Woitaszek2009}, and several minor revisions followed as the 
quality of asteroseismic data from the {\it Kepler} mission gradually improved 
\citep{Mathur2012, Metcalfe2014, Creevey2017}. The first major revision coupled the same 
optimization method to the MESA \citep{Paxton2011} and GYRE \citep{Townsend2013} codes. 
Although most of the input physics in the new version of AMP were chosen to be the same 
as in the previous version, there were two major updates that addressed the dominant 
sources of systematic error in the analysis of {\it Kepler} data sets. First, although 
the Aarhus models included diffusion and settling of helium \citep{MichaudProffitt1993}, 
the treatment of heavier elements was numerically unstable. The MESA models include 
diffusion and settling of both helium and heavier elements \citep{Thoul1994}. Second, the 
original version of AMP included an empirical correction for inadequate modeling of 
near-surface layers \citep{Kjeldsen2008}, while the updated version uses a 
physically-motivated correction that has now become the standard in the field 
\citep{Ball2014}.

The asteroseismic radius, mass, and age of \bethyi from the YREC and MESA modeling 
procedures are listed in Table~\ref{tab2}. The consistency of the asteroseismic 
properties from these two independent model grids and fitting methodologies suggests that 
they are robust at the indicated level of precision.

%%%%%%%%%%%%%%%%%%%%%%%%%%%%%%%%%%%%%%%%%%%%%%%%%%%%%%%%%%%%%%%%%%%%%%%%%%%%%%%%%%%%%%%%%% 
\section{Interpretation}\label{sec4}     

\subsection{Rotational Evolution Modeling}\label{sec4.1}

We performed rotational evolution modeling following a procedure similar to that 
described in \cite{Metcalfe2020}. We used a tracer code that computes the torque on the 
star given the braking law of \cite{vanSaders2016} and the stellar structure as a 
function of age. We utilized two different evolutionary codes to predict the stellar 
structure. The first was a YREC model grid that assumed a chemical enrichment law for 
helium as a function of metallicity but a free mixing length parameter for convection, 
with input physics described in \cite{Metcalfe2020}. The second grid was constructed with 
MESA, tuned to match the input physics of AMP~2.0 \citep{AMP2023} and described in detail 
by \cite{Saunders2024}. This grid allowed both the mixing length and helium to be free 
parameters.

For the braking law, we adopted the form of \cite{vanSaders2016} and the calibration of 
\cite{Saunders2024}, which determined best-fit braking law parameters using a Bayesian 
hierarchical framework to model both individual asteroseismic and open cluster system 
parameters, as well as the parameters of the braking law itself, which were assumed to be 
the same for all stars in the sample. We adopted their critical Rossby number 
$\mathrm{Ro_{crit} = 0.94\pm0.04\ Ro_{\odot}}$ ($\mathrm{Ro_{\odot} = 2.33}$) and braking 
normalization $f_K = 7.64$ for the YREC model grid, and $\mathrm{Ro_{crit} = 0.91\pm0.03\ 
Ro_{\odot}}$ ($\mathrm{Ro_{\odot} = 2.05}$) and $f_K = 5.46\pm0.51$ for the MESA model 
grid. We also adopted their fits of the standard spin-down model to the data (without 
WMB), where $f_{K,YREC}=8.55$ and $f_{K,MESA}=6.11$.

For both model grids, we used $T_{\rm eff}$, [Fe/H], $L_{\rm bol}$, and their 1$\sigma$ 
uncertainties as surface constraints. We adopted a wide Gaussian prior on mass ($\sigma = 
0.5\ M_{\odot}$) from asteroseismology and bulk starting metallicity ($\sigma = 0.3$~dex) 
but a tight asteroseismic prior on age ($6.46 \pm 0.13$~Gyr for the YREC grid, and $6.26 
\pm 0.57$~Gyr for the MESA grid). This was done to help mitigate the ``overcounting'' of 
constraints: the asteroseismic modeling adopted the measured surface properties as 
constraints, so knowledge of them is already reflected in the asteroseismic mass. Age is 
weakly constrained by measurements of $T_{\rm eff}$, [Fe/H], and $L_{\rm bol}$, but very 
tightly constrained by the small frequency separation and the avoided crossing. In this 
sense the asteroseismic age is more independent of the surface constraints than the mass. 
For this reason we adopted the tight prior on the asteroseismic age, but only very broad 
priors on other asteroseismic properties. We assumed a mixing length prior of $1.7 \pm 
0.2$ bounded [1.4, 2.0] for both grids. In the YREC grid, the initial helium 
abundance was treated with a chemical enrichment law; in the MESA grid it was free to 
vary, and assigned a prior of $0.26\pm0.10$, bounded [0.22, 0.28]. We considered 
values below the primordial helium abundance to allow for the possibility of systematic 
errors in the stellar models, and to avoid truncating the posterior probability 
distributions. Note that the measured rotation period from Section~\ref{sec2.2} was not 
used as an input constraint, but only as a consistency check with the predictions from 
two different sets of braking models.

For the YREC standard spin-down model we adopt the best-fit \cite{Saunders2024} braking 
law parameters without uncertainties, and predict a rotation period of $39^{+6}_{-7}$ 
days. For the WMB model, we predict a period of $19.4 \pm 3.5$ days. In the MESA grid, we 
use the neural network framework of \cite{Saunders2024} to predict periods using 
No-U-Turn Sampling \citep{Hoffman2014}, which is a modified Hamiltonian Markov Chain 
Monte Carlo algorithm. The adopted $\mathrm{Ro_{crit}}$ and $f_K$ were treated as 
parameters with tight Gaussian priors given by their quoted uncertainties in 
\cite{Saunders2024}. The standard spin-down model predicts a rotation period of 
$35^{+11}_{-8}$~days, while the WMB model predicts $19^{+5}_{-4}$~days. These values 
are substantially unchanged if we adopt the asteroseismic properties of the lower mass 
model from MESA, with the same wide mass prior now centered at $1.059\ M_\odot$ and a 
tight age prior of $6.10 \pm 0.19$~Gyr.

% FIGURE 5 ---------------------------------------------------------------   
 \begin{figure}[t]
 \centering\includegraphics[width=\columnwidth,trim=7 7 5 0,clip]{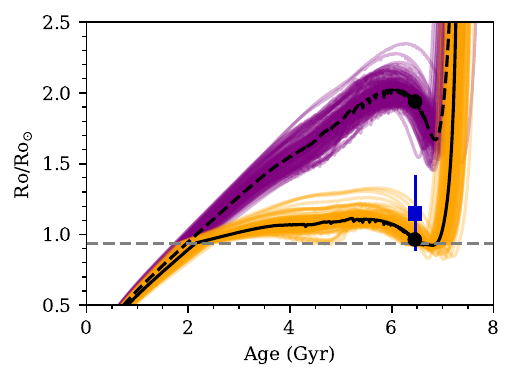}
 \caption{Evolution of the Rossby number for \bethyi in the YREC model grid, with samples 
drawn from the posterior of the standard spin-down model (purple curves) and from the WMB 
model (orange curves). The value of $\mathrm{Ro_{crit}}$ is shown as a gray dashed line, 
while the best-fit value of the Rossby number is shown as a circle for both braking law 
prescriptions. The actual position of \bethyi (assuming the local convective 
turnover time of the best fit model but the measured period and age) is shown as a blue 
square with an error bar.\label{fig5}}
 \end{figure}
%-------------------------------------------------------------------------      

In both grids, despite differing model physics and assumptions about the helium, the WMB 
model predicts a period consistent with that recovered from the current TESS photometry, 
albeit slightly shorter. As found by \cite{Saunders2024}, this is a common pattern in 
more evolved stars, and may arise from our simplified braking prescription. Beyond 
$\mathrm{Ro_{crit}}$ magnetic braking is assumed to cease entirely, but in reality there 
may still be some minimal spin-down that subtly slows the rotation (along with the 
increasing moment of inertia) during the latter half of the main-sequence lifetime. 
Regardless, standard spin-down models predict much longer rotation periods that are 
inconsistent with the observations.

Figure~\ref{fig5} illustrates the evolution of the Rossby number in the YREC model grid 
as a function of stellar age, with the current observed position of \bethyi marked and 
curves for both the standard spin-down (purple) and WMB (orange) cases. Standard 
spin-down models predict a relatively large Rossby number by the age of \bethyi, while 
WMB models predict a Rossby number much closer to the solar value, where stellar activity 
cycles are most prominent \citep{Egeland2017}.

\subsection{The Born-Again Dynamo Phenomenon}\label{sec4.2}

The properties of \bethyi provide a second example of the ``born-again'' dynamo 
phenomenon that was previously suggested to explain the cycle in \aqr 
\citep{Metcalfe2020}. In that scenario, \bethyi started as an F-type star on the 
main-sequence, where it probably had a short-period activity cycle like $\iota$\,Hor 
\citep{Metcalfe2010}. The cycle would gradually grow longer as the rotation rate slowed 
due to magnetic braking over the next $\sim$2~Gyr, until it reached the critical Rossby 
number for the onset of WMB. During the second half of its main-sequence lifetime, the 
rotation rate would remain almost constant while the cycle would grow longer and weaker 
before disappearing entirely \citep{Metcalfe2017}. However, when hydrogen shell-burning 
began and the star expanded and cooled onto the subgiant branch, the longer turnover time 
in the deeper convection zone would overwhelm the longer rotation period from 
conservation of angular momentum. For stars slightly more massive than the Sun, these 
evolutionary effects can push the Rossby number back below $\mathrm{Ro_{crit}}$, 
reinvigorating the large-scale dynamo and brieﬂy sustaining an activity cycle once again 
before ascending the red giant branch. The current position of \bethyi in 
Figure~\ref{fig5} generally supports this evolutionary scenario, which only returns below 
$\mathrm{Ro_{crit}}$ in the WMB scenario.

The activity cycle of \aqr is much more prominent in Ca~HK than the cycle in \bethyi. The 
IUE data for \bethyi show a magnetic maximum in 1986.9, followed by a rise toward the 
subsequent maximum in late 1998 (top panel of Figure~\ref{fig1}). The Ca~HK observations 
from mid-2007 to 2013 span the next predicted maximum in late 2010, and show a slightly 
higher mean activity level during the 2010 season followed by a gradual decline from 
2011--2012 (bottom panel of Figure~\ref{fig1}). Earlier Ca~HK measurements during the 
cycle minimum in 1993 showed a similar range \citep[$S=0.151$--0.171;][]{Henry1996}, so 
the cycle amplitude in the optical appears to be substantially smaller than in the 
ultraviolet. This is a consequence of the larger dynamic range for magnetic variability 
at higher energies, which also explains the clear \mbox{X-ray} cycle in $\alpha$\,Cen\,A 
\citep{Ayres2023} that has not been detected in the optical. These stars might have been 
classified as ``flat activity'' targets from their optical data alone, but the 
availability of higher-energy observations revealed their cycles. By contrast, Mount 
Wilson observations of \aqr show a robust cycle in Ca~HK \citep{Metcalfe2020}.

The difference between the Ca~HK variability of \bethyi and \aqr may be a consequence of 
their relative Rossby numbers. Because rotation and activity are strongly coupled prior 
to the onset of WMB, $\mathrm{Ro_{crit}}$ can be converted into a critical activity level 
\citep[$\log R'_{HK}\!\!\approx\!-4.95$;][]{Brandenburg2017}. For any given star, this 
critical activity level corresponds to a specific Mount Wilson S-index 
($\mathrm{S_{crit}}$), which depends on the B$-$V color \citep{Noyes1984}. For \bethyi 
this value is $\mathrm{S_{crit}}=0.165$, which falls above most of the observations shown 
in the bottom panel of Figure~\ref{fig1}. By contrast, the critical activity level for 
\aqr is $\mathrm{S_{crit}}=0.18$, which falls within the range of variability in the 
Mount Wilson measurements \citep[$S=0.136$--0.196;][]{Metcalfe2020}. Thus, the cycle is 
weaker in \bethyi because it is closer to $\mathrm{Ro_{crit}}$, while \aqr is well below 
$\mathrm{Ro_{crit}}$ and has a stronger cycle.

%%%%%%%%%%%%%%%%%%%%%%%%%%%%%%%%%%%%%%%%%%%%%%%%%%%%%%%%%%%%%%%%%%%%%%%%%%%%%%%%%%%%%%%%%%
\section{Summary and Discussion}\label{sec5}

We have used two months of TESS observations to characterize the solar-type subgiant 
\bethyi and investigate the nature of its activity cycle (Figure~\ref{fig1}). We 
extracted a custom light curve from the data, reducing the noise by a factor of two 
(Section~\ref{sec2.1}) and enabling the first direct measurement of the rotation period 
(Section~\ref{sec2.2}). Analysis of the solar-like oscillations (Section~\ref{sec3.1}) 
identified 23 individual frequencies for detailed asteroseismic modeling 
(Table~\ref{tab1}), yielding precise estimates of the stellar radius, mass, and age 
(Section~\ref{sec3.2}). The resulting stellar properties (Table~\ref{tab2}) provided 
inputs for rotational evolution modeling (Section~\ref{sec4.1}), showing that the 
rotation period of \bethyi is consistent with WMB. In addition, the current Rossby number 
of \bethyi is comparable to \aqr (Figure~\ref{fig5}), which may help explain the 
existence of its activity cycle. We conclude that changes in stellar structure on the 
subgiant branch can reinvigorate the large-scale dynamo and briefly sustain magnetic 
activity cycles (Section~\ref{sec4.2}), a phenomenon that was originally suggested to 
explain the cycle in \aqr \citep{Metcalfe2020} and can best be understood within the WMB 
scenario.

% FIGURE 6 ---------------------------------------------------------------   
 \begin{figure}[t]
 \centering\includegraphics[width=\columnwidth]{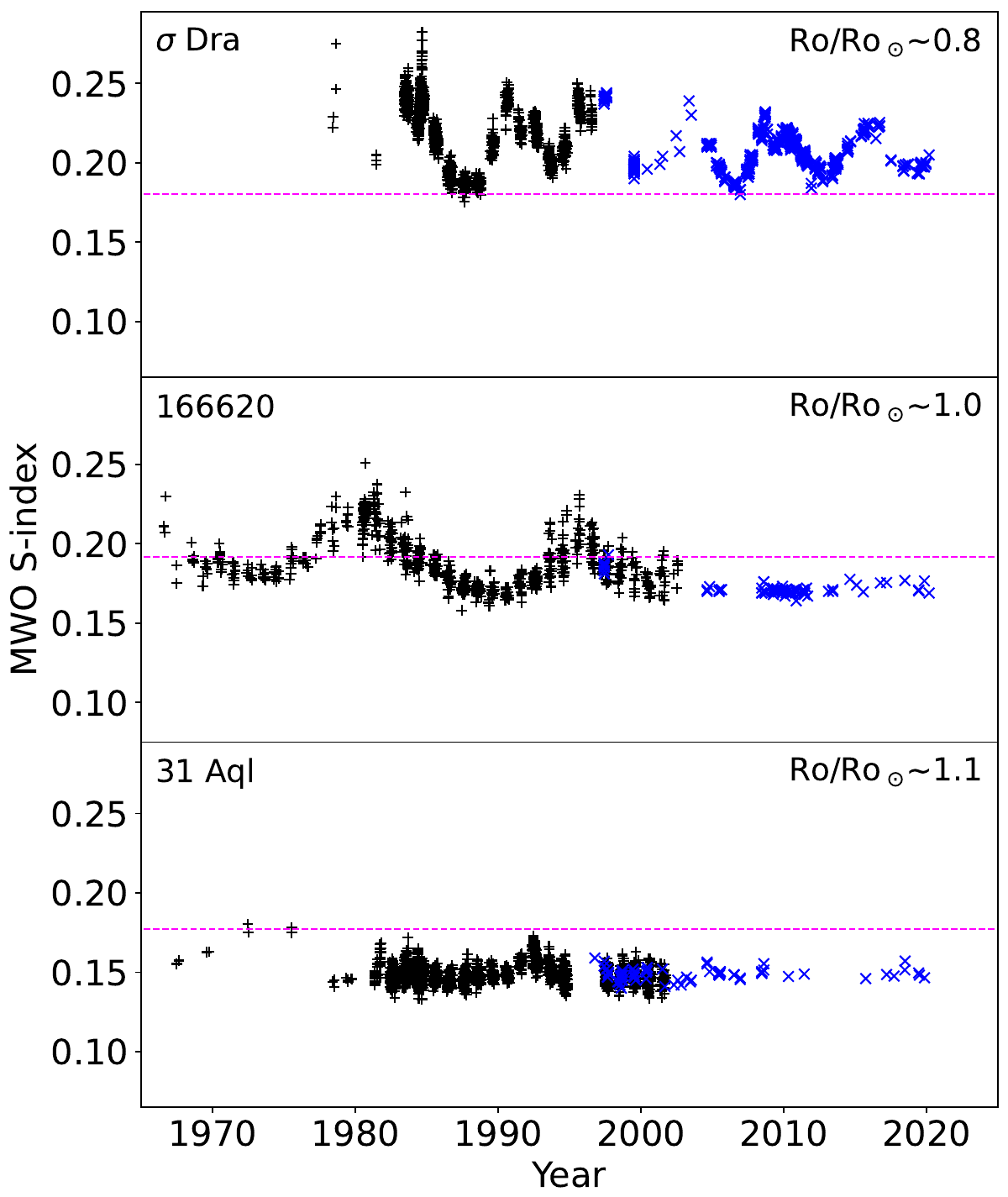}
 \caption{Evolution of stellar activity cycles through the critical activity level that 
corresponds to $\mathrm{Ro_{crit}}$ (magenta dashed lines). Mount Wilson data are shown 
with black plus symbols, while Keck data are shown with blue crosses \citep{Baum2022}. 
Close to the onset of WMB, the critical activity level is near cycle minimum as in 
$\sigma$\,Dra (top). As the mean activity level decreases with age, the cycle can develop 
intermittency as in HD\,166620 (middle). Continued evolution pushes the mean activity 
level permanently below the critical value, and cycles disappear entirely as in 31\,Aql 
(bottom).\label{fig6}}
 \end{figure}
%-------------------------------------------------------------------------               

The critical Rossby number for the onset of WMB ($\mathrm{Ro_{crit}}$) might also 
represent a threshold beyond which large-scale dynamos have difficulty driving cycles 
\citep{Tripathi2021}. As an illustration of the empirical evidence for this idea, the 
long-term activity records from \cite{Baum2022} are shown for three stars in 
Figure~\ref{fig6}. Observations from the Mount Wilson survey are shown with black plus 
symbols, those from Keck are shown with blue crosses, and the critical activity level for 
each star ($\mathrm{S_{crit}}$; see Section~\ref{sec4.2}) is indicated with a magenta 
dashed line. The K0 dwarf $\sigma$\,Dra (top panel) is approaching $\mathrm{Ro_{crit}}$ 
and shows a clear activity cycle with a minimum level comparable to $\mathrm{S_{crit}}$. 
The less active K2 dwarf HD\,166620 (middle panel) initially shows a clear cycle with a 
mean activity level similar to $\mathrm{S_{crit}}$ before the star apparently enters a 
magnetic grand minimum \citep{Luhn2022}. This type of intermittency is predicted to 
become more frequent and more prolonged as the mean activity level continues to decline 
with age \citep{Vashishth2023}. During this phase, active regions that emerge with 
unusual properties (e.g.\ violating Hale's polarity law or Joy's tilt angle law) can 
switch the dynamo between the cycling and non-cycling states \citep{Nagy2017}. 
Eventually, continued evolution pushes the mean activity level so far below 
$\mathrm{S_{crit}}$ that even the most extreme active regions can no longer shift the 
dynamo between these states, and cycles disappear entirely as in the G7 subgiant 31\,Aql 
(bottom panel). Most of the ``flat activity'' stars in the Mount Wilson survey appear to 
be in this permanently low activity regime \citep{Egeland2017}.

Recent direct estimates of the wind braking torque in old solar-type stars have revealed 
an unexpected decline in the large-scale magnetic field as well as the mass-loss rate 
\citep{Metcalfe2021, Metcalfe2022, Metcalfe2023, Metcalfe2024}, and \bethyi can connect 
these effects to underlying changes in the stellar dynamo. We suggest that 
$\mathrm{Ro_{crit}}$ corresponds to a rotation rate that is too slow to imprint 
substantial Coriolis forces on the global convective patterns\footnote{Note that the 
onset of WMB at $\mathrm{Ro_{crit}}$ may prevent most main-sequence stars from reaching 
the higher Rossby numbers above $\sim$1.1~Ro$_\odot$ that are required in global 
convection simulations to generate anti-solar differential rotation \citep{Noraz2024}, so 
stars that rotate faster at the pole and slower at the equator are primarily expected on 
the red giant branch.}. Consequently, related properties such as differential rotation, 
meridional circulation, and tilted active region emergence begin to be disrupted. The 
loss of shear from differential rotation weakens the $\Omega$-effect, inhibiting the 
production of buoyant magnetic loops within the convection zone and yielding shallower 
tilt angles when they ultimately emerge. As the near surface convection gradually shreds 
bipolar magnetic regions, the shallower tilt and weaker differential rotation leads to 
enhanced cancellation of magnetic flux, and the weaker meridional circulation transports 
less of the residual flux toward the polar regions to seed the regeneration of 
large-scale field. This leads to a downward spiral of both flux emergence and the 
production of large-scale magnetic field on stellar evolutionary timescales. With a 
higher fraction of the remaining field concentrated in smaller spatial scales, the 
diminished large-scale field weakens magnetic braking \citep{Reville2015, Garraffo2016, 
See2019} and the increased magnetic complexity throttles the stellar wind from the 
smaller area with open magnetic field lines \citep{Garraffo2015, Shoda2023}, as suggested 
by the recent wind braking estimates.

Future observations of \bethyi will enable a direct estimate of the wind braking torque 
during the ``born-again'' dynamo phase, providing new constraints on the late stages of 
magnetic stellar evolution. Such an estimate will require spectropolarimetry to constrain 
the large-scale magnetic morphology \citep{FinleyMatt2018}, Ly$\alpha$ analysis or an 
\mbox{X-ray} flux to estimate the mass-loss rate \citep{Wood2021}, the rotation period 
determined in Section~\ref{sec2.2}, and the asteroseismic radius and mass determined in 
Section~\ref{sec3.2}. Spectropolarimetry of \bethyi was obtained in July 2024 with 
HARPSpol, and Ly$\alpha$ measurements have been approved for the Hubble Space Telescope, 
so we should soon learn how the resurgence of a stellar activity cycle affects the 
current rate of angular momentum loss in this evolved subgiant.

%%%%%%%%%%%%%%%%%%%%%%%%%%%%%%%%%%%%%%%%%%%%%%%%%%%%%%%%%%%%%%%%%%%%%%%%%%%%%%%%%%%%%%%%%% 
\vspace*{12pt}
We would like to thank Axel Brandenburg, Paul Charbonneau, and Ricky Egeland for helpful discussions.
This paper includes data collected with the TESS mission, obtained from the Mikulski Archive for Space Telescopes at the Space Telescope Science Institute (STScI). The specific observations analyzed can be accessed via \dataset[doi:10.17909/g809-wf30]{https://doi.org/10.17909/g809-wf30}. Funding for the TESS mission is provided by the NASA Explorer Program. STScI is operated by the Association of Universities for Research in Astronomy, Inc., under NASA contract NAS 5–26555.
T.S.M.\ acknowledges support from NASA grant 80NSSC22K0475. Computational time at the Texas Advanced Computing Center was provided through XSEDE allocation TG-AST090107.
D.H.\ acknowledges support from the Alfred P. Sloan Foundation, NASA (80NSSC22K0303, 80NSSC23K0434, 80NSSC23K0435), and the Australian Research Council (FT200100871).
R.A.G.\ acknowledges support from the PLATO and SoHO/GOLF grants of the Centre National D'{\'{E}}tudes Spatiales.
S.N.B.\ acknowledges support from PLATO ASI-INAF agreement no.\ 2022-28-HH.0 ``PLATO Fase D''.
M.B.N.\ acknowledges support from the UK Space Agency.
J.M.J.O.\ acknowledges support from the NASA Hubble Fellowship grant HST-HF2-51517.001, awarded by STScI.
N.S.\ acknowledges support from the NSF Graduate Research Fellowship Program under grant numbers 1842402 \& 2236415.
This research benefited from discussions at the workshop ``Solar and Stellar Dynamos: A New Era'', hosted and supported by the International Space Science Institute in Bern, Switzerland.

%%%%%%%%%%%%%%%%%%%%%%%%%%%%%%%%%%%%%%%%%%%%%%%%%%%%%%%%%%%%%%%%%%%%%%%%%%%%%%%%%%%%%%%%%% 
%\bibliographystyle{aasjournal}
%\bibliography{references}

\end{document}